\documentclass[a4paper]{article}

\usepackage{INTERSPEECH2021}
\usepackage{tabularx}
\usepackage{amsmath,amsfonts,amssymb,amsthm}
\usepackage{tikz}
\usepackage{url}
\renewcommand{\v}[1]{\mathbf{#1}}

\title{Probing neural audio codecs for distinctions among English nuclear tunes}
\name{Juan Pablo Vigneaux and Jennifer Cole}
\address{
  Northwestern University}
\email{$\{$jpvigneaux, jennifer.cole1$\}$@northwestern.edu}

\begin{document}

\maketitle

\begin{abstract}
     State-of-the-art spoken dialogue models (Défossez et al. 2024; Schalkwyk et al. 2025) use neural audio codecs to ``tokenize'' audio signals into a lower-frequency stream of vectorial latent representations, each quantized using a hierarchy of vector codebooks. A transformer layer allows these representations to reflect some time- and context-dependent patterns. We train probes on labeled audio data from Cole et al. (2023) to test whether the pitch trajectories that characterize English phrase-final (nuclear) intonational tunes are among these patterns. Results: Linear probes trained on the unquantized latents or some of the associated codewords yield above-chance accuracy in distinguishing eight phonologically specified nuclear tunes with monotonal pitch accents (top average test accuracy (TATA): 0.31) and the five clusters of these tunes that are robust in human speech production and perception (TATA: 0.45). Greater accuracy (TATAs: 0.74-0.89) is attained for binary distinctions between classes of rising vs. falling tunes, respectively used for questions and assertions. Information about tunes is spread among all codebooks, which calls into question a distinction between ‘semantic’ and ‘acoustic’ codebooks found in the literature. Accuracies improve with nonlinear probes, but discrimination among the five clusters remains far from human performance, suggesting a fundamental limitation of current codecs. 
\end{abstract}

\section{Introduction}

\subsection{Motivation} 

Motivated by the goal of understanding the prosodic capabilities and limitations of two state-of-the-art spoken dialogue models \cite{Defossez2024,Schalkwyk2025}, we conducted a series of experiments to assess the detectability of pitch patterns ('tunes') in the compressed representations of audio on which these conversational systems operate. In English, as in many other languages, phrasal pitch patterns convey the speaker's communicative intentions (speech acts such asking or telling) and the contribution of individual words and phrases to the evolving information landscape of the discourse  \cite{prieto_intonational_2015,westera_meanings_2020}. Given the critical role of intonation for encoding these kinds of pragmatic meaning, a successful spoken dialogue model must be able to dissociate intonation patterns from the phone and word content of speech, to approximate the communication behavior of human speakers.  

The spoken dialogue models we examine rely on a first stage of audio ``quantization'' performed by a neural audio codec, Mimi. Mimi has the structure of an autoencoder with a discrete bottleneck and is trained  on masked audio reconstruction, while aligning via distillation one of its discrete codebooks to the latent representations of WavLM \cite{chen_wavlm_2022}, which is regarded as 'semantic' (or more accurately, phonological) in nature \cite{Defossez2024,Schalkwyk2025}.  
Our question is whether intonation is represented in this 'semantic', WavLM-aligned quantized encoding, or in other discrete codebooks of the codec. For this, we use probing experiments to evaluate differences in unquantized and quantized Mimi-generated latent representations of recorded utterances from American English speakers that represent eight intonational tunes of American English. 

\subsection{Probes and interpretability}

Probes were first introduced in \cite{Alain2018} to understand the inner workings of deep neural networks. The idea is to evaluate how well the representations of the input in intermediate layers serve for solving an auxiliary classification problem. 
The auxiliary classification problem can be viewed as a test for the presence of a ``latent feature'' that is judged relevant to the original goal of the DNN; in this form, linear probes are a fundamental tool in \emph{mechanistic interpretability} \cite{Sharkey2025}.  For instance, in \cite{Nanda2023a,Li2023} probes are used to assess the encoding of board states in the intermediate layers of a transformer trained on predicting next moves in the game Othello, a simplified version of chess.



\subsection{Nuclear tunes}\label{sec:intro_tunes}

We focus here on the possibility of training probes that classify latent representations of audio recordings of utterances generated by Mimi in terms of English phrase-final nuclear intonational tunes they implement, or clusters thereof. There is an eight-way distinction in the pitch trajectories of nuclear tunes that follows from combining High and Low tones from the (monotonal) pitch accents ($H^*, \,L^*$), phrase accents ($H-$,\, $L-$), and boundary tones ($H\%,\, L\%$) as characterized in the phonological Autosegmental-Metrical model of American English intonation \cite{ladd_intonational_2008}. Hereafter we drop the diacritics and refer  to these \textit{eight basic tunes} in their three components, e.g. \textit{hlh}. 

Cole at al. \cite{Cole2023} tested the status of these eight tunes as distinct from one another in speech production using a tune imitation task, and in speech perception using an AX discrimination task, with native speakers of American English. The results, consistent across modalities, reveal robust distinctions among five tune classes, which overlap almost perfectly with the following partition of the imitations in terms of their target tune: $\{lll\}$, $\{llh,lhl\}$, $\{lhh\}$, $\{hll, hlh\}$, and $\{hhh, hhl\}$; the only exception is that $17\%$ of the imitations of $hhh$ and $hhl$ fall in the cluster that contains almost all the $lhh$ imitations. For details, see  \cite{Cole2023}, specially Figure 5. We take the neat clustering analysis as evidence that at least these five tune classes have distinct representations in the minds of these speakers.

Among the eight nuclear tunes tested in this study, the initial tone encodes information structure distinctions (focus, givenness), while the later two tones encode speech act distinctions (question, assertion, listing); hence, tunes  are  crucial for reasoning about pragmatic meaning in spoken English \cite{prieto_intonational_2015, westera_meanings_2020}.  We therefore postulate that a good `semantic encoding' of audio data|which would serve as a useful compression of the audio for meaningful, context-aware next-token prediction|should include a speaker-independent encoding of these tunes. 

\section{Methods}

\subsection{Labeled audio} 

We work with audio data from \cite{Cole2023}, consisting  of 4656 audio recordings of utterances, each consisting of a short sentence produced with one of the eight basic nuclear tunes. These audio recordings have two origins. 48 are pitch-resynthesized audio recordings from base recordings of one male and one female speaker (8 tunes x 3 model sentences x 2 speakers). These resynthesized 
signals comprised the stimuli for a tune imitation task where participants (\textit{N} = 30) imitated the tune from an audio stimulus in the read-aloud production of a new sentence. 4608 audio recordings are these imitated productions. These recordings constitute the best available non-synthetic data that corresponds systematically to the eight phonologically specified nuclear tunes of American English tested here. The resynthesized utterances were produced with three model sentences: ``Her name is Marilyn,'' ``She quoted Helena,'' ``He answered Jeremy,'' while the imitated utterances were produced with three different sentences: ``She remained with Madelyn,'' ``They honored Melanie,'' ``He modeled Harmony.'' In all sentences, the final word 
carried the nuclear tune. The intonation over the pre-nuclear region of the sentences was held constant across the pitch-resynthesized recordings, and was not analyzed in the imitated productions. Audio files were aligned with phone and word content using the Montreal Forced Aligner \cite{MFA}, to locate the time points that delimit the final, accented word.

In terms of classification tests, we are not only interested in assessing distinctions between the eight nuclear tunes or the five emergent clusters over the imitations of those tunes, as described in Section \ref{sec:intro_tunes}. Some simpler binary distinctions are also of interest, particularly in connection with pragmatic reasoning. The classification problems we consider are summarized in Table \ref{tab:classifications}; this is not an exhaustive list of all the pragmatically-relevant distinctions, but it  captures the most fundamental ones. 

\begin{table}
\caption{Classification problems considered in this work.}\label{tab:classifications}
\centering
\begin{tabularx}{\columnwidth}{>{\hsize=.5\hsize\linewidth=\hsize}X
>{\hsize=1.5\hsize\linewidth=\hsize}X}
\toprule
\textbf{Classification problem} & \textbf{Description} \\\midrule
\textsf{8class} & The eight nuclear tune shapes that follow from combining (monotonal) pitch accents, phrase accents, and boundary tones in the Autosegmental-Metrical model. \\
\hline
\textsf{5class} & The five clusters of nuclear tunes obtained in \cite{Cole2023}, as described in Section \ref{sec:intro_tunes}.   \\
\hline
\textsf{hhh-vs-lll} & The greatest magnitude distinction: high-rising vs. low-falling. \\
\hline
\textsf{hxx-vs-lxx} & A distinction only in pitch accent \\
\hline
\textsf{xll-vs-xhh} & A distinction only in edge tones: falling (xll) vs. rising (xhh), underlying the distinction between assertion and question. \\
\hline
\bottomrule
\end{tabularx}
\end{table}
 
For each classification problem, we split the labeled data into train, development (dev) and test sets, respectively $70\%$, $15\%$ and $15\%$ of the total amount of audio samples relevant for the problem. We maintain the original proportion of tune labels in each set.

\subsection{Neural encodings}

We work with the Mimi neural audio codec \cite{Defossez2024}, which consists of three components: an encoder, a quantizer, and a decoder. 

The encoder projects a 24 kHz single-channel waveform to a 12.5 Hz sequence of 512-dimensional latent representation (below we refer to these as \emph{unquantized} latent representations) by applying first a convolutional  network, followed by a transformer block. The transformer block has a finite context of 250 frames (20 seconds). Convolutions are causal, so that Mimi can run in a streaming regime (real time). 

The quantizer applies, in parallel, vector quantization (VQ) and residual vector quantization (RVQ) to the unquantized latent representations. Vector quantization is associated to a single finite codebook (which we call \emph{Codebook 0}), made of 256 dimensional vectors (its \emph{codewords}), used to directly approximate the unquantized latent representation. These codewords are aligned, via distillation, with the discrete latent representations generated by WavLM \cite{chen_wavlm_2022}, a foundation speech-model pretrained on masked  speech prediction and audio denoising; since WavLM representations are useful for a wide range of speech-analysis tasks, but not so for audio generation, they are regarded  as `semantic' \cite{Defossez2024}.\footnote{Unlike Mimi, WavLM is \emph{not} a causal model, hence it is unsuitable for the streaming regime.}    In turn, residual vector quantization yields a hierarchy of seven finite codebooks, called here \emph{Codebook 1}, ...,  \emph{Codebook 7}. Only the first of these is meant to  approximate the unquantized latent representation: the second approximates the defect in this first order-approximation, the third the defect in the second-order approximation, and so on. The RVQ codewords are also 256-dimensional.

The decoder part mirrors the encoder. It is pre-trained to provide good audio reconstruction quality even if it only has access to the quantized encodings. For this reason, the spoken dialogue models in \cite{Defossez2024,Schalkwyk2025} consist of a couple of transformers that predict these codewords autoregressively.

\subsection{Probes}

Our probes consist of three consecutive steps: aggregation of the latent representations through a weighted average, dimensionality reduction of the resulting vector through PCA, and training of a linear classifier that assigns tune class labels to the PCA-reduced, aggregated latent representations. 


\subsubsection{Aggregation of the latent representations}

For each sample, we leverage the time stamps in the TextGrid to extract the sequence $(\v x_0,...,\v x_{N})$ of latent representations that correspond to the accented word (the final word). These vectors $\v x$ stand for the unquantized latent representations or codewords.  We aggregate the vector by means of a weighted average,
$ \v y = \sum_{i=0}^{N} w_i \v x_i.$
The weights $w_i$ depend on two hyperparameters of our model, $\delta_F$ and $\delta_B$, that we call, respectively, the \emph{forward decay rate} and the \textit{backward decay rate}. When $\delta_F=\delta_B= 0$, we recover a standard average; such averages are used frequently in the probing literature to aggregate vectorial representations. We allow here more flexibility, though: if $\delta_B \gg \delta_F$, then most of the weight is assigned to the final portion of the representations (corresponding to the intervals of the phrase accent and boundary tone). We also allow the symmetric case $\delta_F\gg\delta_B$, which could  be relevant when trying to distinguish classes defined primarily by the pitch accent.\footnote{Explicitly, 
\begin{equation}
  w_i = \frac{\exp(-\delta_F\operatorname{d}(0,i) - \delta_B\operatorname{d}(i,N))}{\sum_{i=0}^N \exp(-\delta_F\operatorname{d}(0,i) - \delta_B\operatorname{d}(i,N))}.
\end{equation}
where  $\operatorname{d}(i,j)$ denotes the distance $|j-i|$ between index $i$ and index $j$.}

\subsubsection{Dimensionality reduction}

We use the training data, assembled in a matrix $Y = \begin{bmatrix}
    \v y^{(1)} & \cdots & \v y^{(N)}\\
\end{bmatrix}^T $ of size $(\text{number of samples}\times  d)$, to perform Principal Component Analysis (PCA), thus obtaining an orthogonal basis $\mathcal B = (\v b_1, ..., \v b_{d})$ of the latent space, given by the principal directions, which are the eigenvectors of $Y^TY$. Above $d$ is the dimension of the representations: $d=512$ in the unquantized case and $d=256$ for the codebooks. 

Once the basis $\mathcal B$ is computed, it is used to reduce any aggregated vector $\v y^{(i)}$|coming from the training, development or test set|by projecting it onto the linear space generated by the first $d_{PCA}$ elements of $\mathcal B$, thus obtaining a vector $\tilde {\v y}^{(i)}$. The integer $d_{PCA}$, which we call the \emph{PCA dimension}, is a hyperparameter of our model. 

Dimensionality reduction is important to reduce the number of learnable parameters and reduce the risk of overfitting or memorization, see discussion in \cite{Alain2018} and also \cite{zhang2017understanding}, where it is shown that large models can fit random labels. 

\subsubsection{Classifiers}
Let $d_c$ denote the number of classes in the classification problem at hand, cf. Table \ref{tab:classifications}. In the context of this work, a \emph{linear classifier} is a function 
\begin{equation}
    f_L(\v y) = \sigma( W \v y + b ) 
\end{equation}
 for some choice of real matrix  $W$ of size $d_c \times d$ and $d_c$-dimensional (bias) vector  $b$. The symbol $\sigma$ denotes the softmax function, which transforms a vector $(z_1,...,z_{d_c})$ into a probability vector with components proportional to $(\exp(z_1),...,\exp(z_{d_c})).$

Similarly, for us,  a \emph{nonlinear classifier} is a function 
\begin{equation}
    f_{NL}(\v y) = \sigma( W_2 \rho( W_1 \v y + b_1) + b_2 ) 
\end{equation}
for some choice of real matrices $W_1$ and $W_2$ of respective sizes $d_c \times d_h$ and $d_h\times d$, and real valued vectors  $b_1$ and $b_2$ of suitable dimensions. Here $\sigma$ denotes again the sigmoid, and $\rho$ consists in applying first Layer Normalization \cite{ba_layer_2016} (an approximate projection on the unit sphere \cite{winsor_re-examining_2022}) followed by the leaky ReLU function with parameter 0.01 \cite{Xu2020a}.
In the expressions above, $d_h$ is the \emph{hidden dimension}, another hyperparameter of our model. 

Both $f_L$ and $f_{NL}$ act on aggregated and dimensionally reduced embeddings $\tilde {\v y}^{(i)}$, for $i=1,...,N$.

\subsection{Details of training}

We train our models to minimize the cross entropy loss. For this, we use the Adam optimizer. Besides the forward decay rate, the backward decay rate, and the PCA dimension, our hyperparameters are the batch size, the learning rate, and the number of training epochs. We start by using \emph{Optuna} \cite{akiba_optuna_2019} to select optimal hyperparameters for each type of probe (linear, nonlinear) on each type of classification problem in Table \ref{tab:classifications}, always working with the unquantized embeddings. We use the hyperparameters thus determined to train the same probes on the codewords (vectors) corresponding to the unquantized representations in each discrete codebook.


\begin{figure}[h!]
  \centering
  \includegraphics[width=\linewidth]{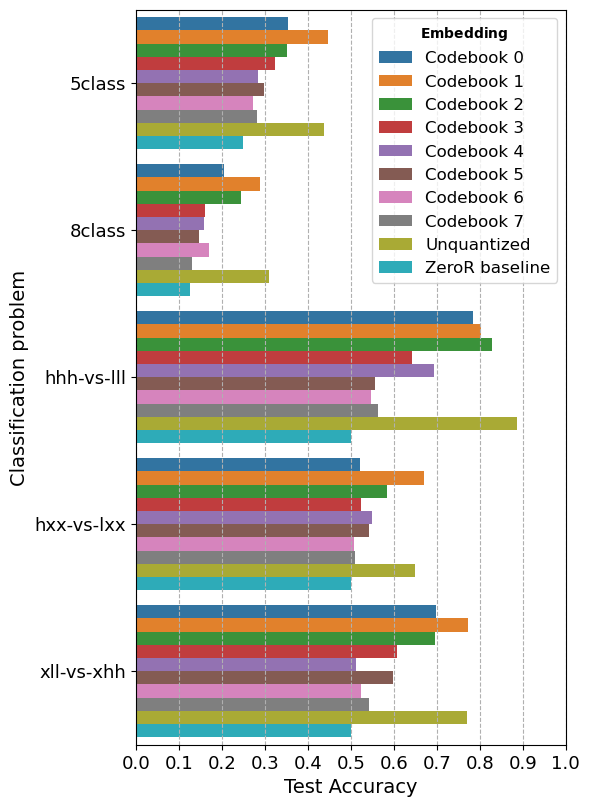}
  \caption{Accuracy of optimal linear probes on their test sets. The colors indicate the kind of input used to train the linear probe. We also represent the ZeroR baseline.}\label{tab:accs_linear}
\end{figure} 

\section{Results}

For any probe, the \emph{accuracy} is the proportion of correctly classified samples. Figure \ref{tab:accs_linear} summarizes the average accuracies obtained by linear probes trained on the classification on the corresponding test sets, together with the ZeroR baseline,  which is the performance of a classifier that assigns to any input the majority class (assuming an exact stratified splitting into train, development and test sets). In each case, these average accuracies are obtained from three independently trained linear probes (with the same optimal hyperparmeters).

We observe that probes trained on unquantized embeddings or their representation in the first three codebooks improve the ZeroR baseline at least $33\%$ (\textsf{hxx-vs-lxx}) and even $126\%$ (\textsf{8class}). Moreover, unquantized embeddings always perform better or as well as probes trained on the codewords, and  in all cases the representation given by Codebook 1 is better than the representation given by Codebook 0 at predicting  tune classes. 

Note that linear probes are less sensitive to the pitch accent than to the edge tune, despite having the freedom to weight the initial samples more strongly during the hyperparameter optimization phase.

\begin{figure}
  \centering
  \includegraphics[width=0.95\linewidth]{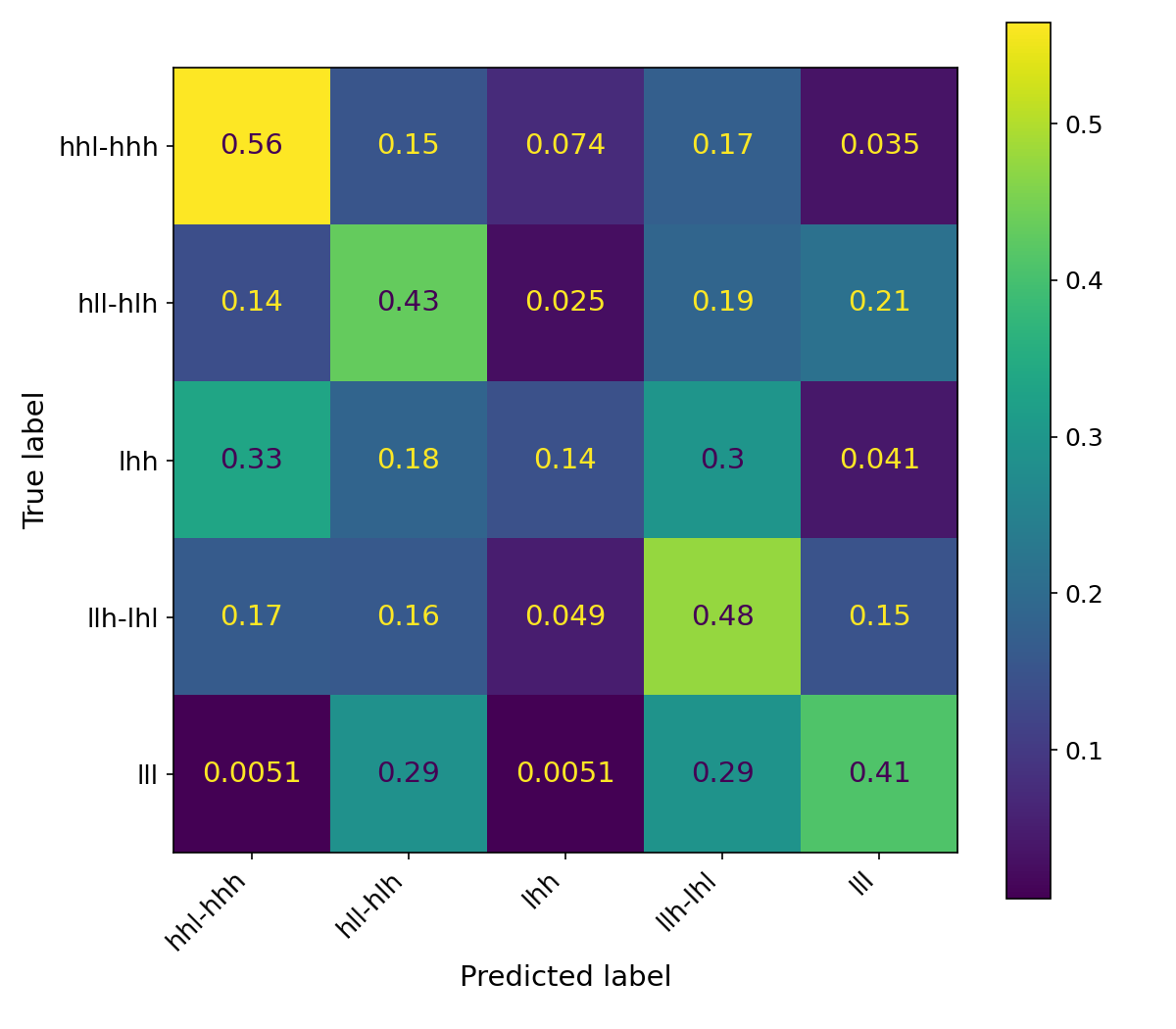}
  \caption{Confusion matrix of the test set predictions generated by the linear probe trained on unquantized embeddings for the 5 class classification problem.}
  \label{fig:5class-cm}
\end{figure}

Figure \ref{fig:5class-cm} depicts the confusion matrix of the trained linear probes on the \textsf{5class} classification task (percentages are averages over the three independently trained probes). It shows that these probes are significantly better than chance at predicting four of the five clusters identified in \cite{Cole2023}, despite introducing much more overlap between the clusters than humans. However, poor performance occurs for $lhh$; recall that already in the clustering of human imitations with respect to target tunes, the cluster that contained $lhh$ also contained $17\%$ of the $hhh$ and $hhl$ tunes, a `confusion' that prominently shows up here. There is also an undesirable overlap between $lhh$ and $\{llh,lhl\}$.

\begin{figure}
  \centering
  \includegraphics[width=\linewidth]{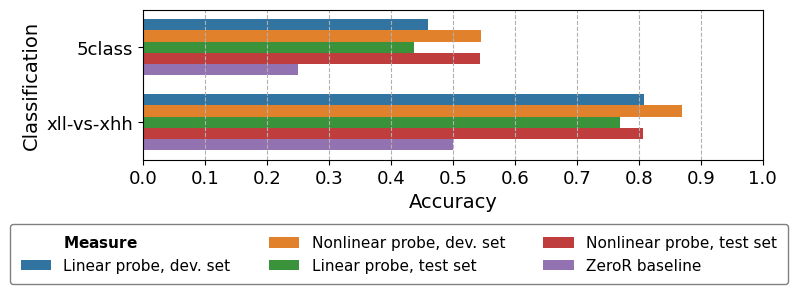}
  \caption{Accuracy of linear and nonlinear probes on unquantized embeddings.}\label{fig:linear-vs-nonlinear}
\end{figure}

For the binary classifications in Table \ref{tab:classifications}, nonlinear probes trained on unquantized embeddings perform just slightly better than the linear ones; the greatest improvement occurs for \textsf{xll-vs-xhh} ($4.8\%$ improvement in the test set). The advantage of nonlinear probes is more noticeable for multi-class classification, particularly for the \textsf{5class} problem ($24.4\%$  improvement in the test set).  While remarkable, the performance is still far from human level. Both cases are illustrated in Figure \ref{fig:linear-vs-nonlinear}.

\section{Discussion }

 The highest accuracy is obtained for the binary classification of \textsf{hhh-vs-lll}. The F0 trajectories are most distinct from one another for this pair. The ~90\% accuracy for the unquantized audio in this task show that Mimi, and hence systems built on top of it, are capable of high-fidelity representation of at least a coarse-grain F0 distinction.

Classification accuracy is higher for the pair that differ in the last two tones (the "edge tones"), \textsf{xll-vs-xhh} compared to the pair that differ only in the initial tone (the pitch accent). Note that due to accentual lengthening, the pitch accent interval comprises nearly half of the overall length of the tune, so the better performance on classifying based on edge tones cannot be due to a length advantage. This might imply that the Mimi-based spoken dialogue models (and similar ones) are less precise in the manipulation of pragmatic distinctions that involve pitch accent such as focus or givenness, but this conclusion is currently difficult to confirm  given the absence of benchmarks that involve pragmatic reasoning. 

Classification accuracy is higher for the \textsf{5class} classification compared to the \textsf{8class}, which accords with the findings from \cite{Cole2023}, where regression and clustering analyses show robust distinctions among the 5 emergent classes, in which two tune pairs merge from the original set of eight tunes tested. 

Importantly, the information about intonational tunes is present in several codebooks, particularly the initial ones (0-2). A similar 'delocalization' of pitch information appears in \cite{Sadok2025}, which analyses the discrete latent representations given by SpeechTokenizer \cite{zhang_speechtokenizer_2024}, a system analogous to Mimi used to discretize audio signals.  As we explained in the introduction, intonational tunes are systematic, speaker-independent units that should play an important role in the generation of context-aware speech. Therefore, we would expect to see them primarily represented in Codebook 0 if this  were indeed  'semantic' in nature, but it is not the case. Instead, the information about tunes appears more prominently in Codebook 1, which is in charge of audio reconstruction (and only to a lesser extent in those codebooks that hierarchically refine its encoding). 

A more detailed analysis of the literature showed us that WavLM, which serves as reference for the distillation of Codebook 0, was not evaluated on any task that involves prosodically-informed inference. In fact, it focused on the SUPERB benchmark \cite{yang_superb_2021}, in which paralinguistic aspects (including prosody) only appear in connection with a single task: emotion recognition. Moreover, performance in the emotion recognition task is highly correlated with the performance in Speaker Identification (SID) and Automatic Speaker Verification (ASV) \cite[Sec. 5.2]{tsai_superb-sg_2022}. Hence SUPERB has no task that favors an speaker-independent representation of prosody, particularly in connection with pragmatic meaning.


\section{Conclusion and Perspectives}

Mimi, a state of the art neural audio codec, generates latent representations that encode, with various degrees of accuracy, fundamental distinctions between intonational tunes. Basic distinctions between high and low pitch, and between rising and falling tunes are encoded with very high accuracy. The detection of pitch accent is less accurate. Similarly, there seems to be a nontrivial distinction between the five clusters of tunes established in \cite{Cole2023}, although it is not as accurate as the distinctions in human perception and production. We wonder how represented these five clusters were in Mimi's training data.

We see a pressing need of extending the existing benchmarks used for evaluating  pre-trained self-supervised speech foundation models (such as SUPERB \cite{yang_superb_2021} and SUPERB-SG \cite{tsai_superb-sg_2022}) and spoken dialogue models (cf. \cite{Schalkwyk2025}), by including pragmatic inferences such as implicatures and speech act resolution that depend on intonational tunes.

Both pretraining and fine tuning can benefit from enriching the training data and extending the learning tasks so that intonation plays a more prominent role. For instance, by using dialogic training materials and promoting learning of prosody-based pragmatic inference (e.g. dialogue act classification  in a supervised setting; next utterance  prediction in a self-supervised setting).    We think this would elicit latent representations that are more sensitive to the pitch accent and the five clusters mentioned above, and in turn enable conversational models capable of finer pragmatic reasoning. 
\\
\\
\noindent\textbf{Acknowledgments.} Project supported by NSF BCS-1944773 to JC.

\bibliographystyle{IEEEtran}
\bibliography{mybib, mybibliography}

\end{document}